\documentclass[twocolumn]{article}
\usepackage{amsmath}
\usepackage{amssymb}
\usepackage{authblk}
\usepackage[parfill]{parskip}
\usepackage{graphicx}
\usepackage{array}
\usepackage{float}
\restylefloat{table}
\usepackage{sectsty}
\sectionfont{\small}
\usepackage{xcolor}
\usepackage{lipsum}
\usepackage{caption}
\usepackage[labelsep=period]{caption}
\usepackage[symbol]{footmisc}

\newcommand{\RNum}[1]{\uppercase\expandafter{\romannumeral #1\relax}}
\usepackage{abstract}
\title {\bfseries Another Look at Flavor}

\author[1,2,3]{W. W. Buck}
\author[2,4]{Stinson Lee}

\affil[1]{\small{School of STEM, University of Washington, Bothell, Washingtion 98011, USA 
		}}
\affil[2]{Department of Physics, University of Washington, Seattle, Washington 98195, USA}
\affil[3]{Department of Physics, College of William and Mary, Williamsburg, Virgina 23187, USA}
\affil[4]{Department of Physics, CCAS, George Washington University, Washington, District of Columbia 20052, USA}

\date{Tuesday, October 16, 2018}

\begin{document}

	\twocolumn[
	\begin{@twocolumnfalse}
		\maketitle
		
		\begin{center}
			\noindent
			A new suggestion for organizing the charged pseudoscalar mesons is presented. The simple NR potential model employs a single value of the light quark mass and determines the potential coupling constant directly from the charged pseudoscalar meson’s, Pion and Kaon, experimental radii values. All other constituent quark masses are directly dependent on the light quark mass value. The model compares features of the traditional approaches of the Non-relativistic(NR) quark model as well as that of a Relativistic quark model. We explore the possibility of introducing flavor as a dynamic quantum number.			
		\end{center} 
	\vspace{2\baselineskip}
	\end{@twocolumnfalse}
	] 	
    
    The discovery of QCD hadrons has revolutionized particle physics; and, has led researchers to better understand the electro-weak-strong force. Organizing mesons according to $J^{PC}$ has become, with great benefit and success, the standard designation and organizing tool of the Standard Model. It is also very clear that the Non-relativistic (NR) quark model has strong calculation strength that provides a good picture of many of the hadronic spectral qualities and quantities[1]. We use that success and consider adding another perspective.  Instead of an ambitious calculation of all the mesons, we choose to examine only the $\pi^{\pm}$, $K^{\pm}$, $D^{\pm}$, and $D^{\pm}$ mesons. We then ask, what would be the impact of this same pseudoscalar meson spectroscopy if a dynamical flavor numerical value, meeting the boundary conditions, were to be introduced. To address this question, we use a simple 1/r potential with no angular momentum nor spin; and thus no spin orbit terms nor relativistic effects either.  We consider this present research direction an exploratory calculation.
    
    For the purpose of examining only the charged pseudoscalars, we consider the up and down quarks to have, as usual, the same mass and designate them light quarks as is typical. And so the Pion is made up of two light quarks and is designated as unflavored.  The other mesons have two unequal mass quarks where one is a light quark.  We designate these charge symmetric mesons, $K^{+}$, $D^{+}$, $B^{+}$, and, eventually the $T^{+}$ (once confirmed), as the usual lowest lying flavor mesons. 
    
    Our simple model approach has the following general basic assumptions: 
    \begin{enumerate}
    	\item Use the lowest order non-relativistic wave equation,
    	\item the light quark mass has the exact same value in all mesons, and
    	\item employ eigenvalues and eigenfunctions of a “Hydrogen-like” wave equation using a 1/r potential, V, with a phenomenological coupling constant.
    \end{enumerate}
	
	As we are exploring the possibility of introducing an additional and new representation of the flavor quantum number, we do not apologize for not having a confining potential in this phase of the research. Even though it is clear that dynamics changes with an additional confining potential, we find we can obtain good values of the charge radii with a simple 1/r potential that begs a more sophisticated potential approach later. As one might expect, the calculated decay constants of these mesons, without the benefit of a confining interaction, are far larger than experimental values. The charge radius is, of course, the slope of the electromagnetic elastic form factor at $Q^2=0$.  Though all of our model calculations are performed in configuration space. This paper reports on our findings.
	
	The three dimensional spherical potential is of the form 
	\begin{equation}
		V(r)=-\frac{\xi^2}{r},
		\label{1}
	\end{equation}
	with $\xi^2$ as a phenomenological coupling strength to be determined.
	
	The Schr\"odinger Equation, to lowest order that does not include spin nor angular momentum to define the odd parity of these mesons, is then solved for two separate Cases: 
	\begin{enumerate}
	 	\item  $n_r =1$  (a revisiting of the non-relativistic quark model [1]); and 
	 	\item  $n_r>1$,    	
	\end{enumerate}
	where $n_r$ is radial quantum number. In both cases, the square of the charge radii are defined as
	\begin{equation}
		<r^2>_{n_r}=\int |\psi|^2_{n_r} r^2 d^3r,
		\label{2}
	\end{equation}
	where $n_r$ represents the given meson and $\psi_{n_r}$ is the corresponding meson eigenfunction.
	
	We first present Case 1($n_r=1$), which essentially reviews aspects of the non-relativistic quark model. In the Case 1 the radial quantum number($n_r$) is 1; and all the charged pseudoscalar mesons listed above will be examined. We begin with the Pion to obtain a starting point for computing charge radii and other observables.
	
	Consider, in natural units, the Pion mass to be
	\begin{equation}
		M=m_{q1}+m_{q2}-\frac{E_0}{n^2_r},
		\label{3}
	\end{equation}
	where $m_{q1}= m_{q2}=m_L=$the light quark(up or down) mass. The radial quantum number $n=n_r=1$ for Case 1. For the equal quark mass meson such as the Pion, we choose $M = M_\pi = 140 MeV$ and obtain $E_0= \frac{\mu \xi^4}{2}$, where $E_0=n^2_r $ times the binding energy; with $\mu$ as the reduced mass. For the equal mass case, the reduced mass is, of course, $\mu=\frac{m_L}{2}$ for the Pion. There are two parameters in Eqn (\ref{3}), $\xi$ and $m_L$. We shall make an assumption about $m_L$ so that an easy solution can be obtained for $E_0$ and thus find $\xi$.  Somewhat arbitrarily, we choose $m_L$ to be $400 MeV$. We note that employing a different value for $m_L$ simply produces a different value for $\xi^2$, a phenomenological constant, equivalent to a slope change that does not affect our major qualitative conclusions.  The quark mass could be any number of values of course; yet we pick a constituent quark mass value not inconsistent with many potential models[2]. Again, choosing another value for $m_L$ will produce a corresponding different value for $\xi^2$ (the coupling constant and slope of the attractive 1/r potential).  As will be seen below, these two parameters are inversely proportional and consistent with Eqn.(\ref{3}) to produce the square charge radius. Again, the completely analytical work reported on here is explorational.  
	
	For $n_r=1$, the Pion wave function, $\psi_{n_r}$, is
	\begin{equation}
		\psi_{1}(r)=\frac{1}{(\pi a^3_0)^{\frac{3}{2}}}e^{-\frac{r}{a_0}}
		\label{4}
	\end{equation}
	with $a_0=\frac{1}{\mu \xi^2}$. Calculating the results for Eqn.(\ref{2}) yields
	\begin{equation}
		<r^2>_\pi=\frac{116427}{\mu^2 \xi^4}fm^2.
		\label{5}
	\end{equation}
	Since $\mu=\frac{m_L}{2}$ and $m_L=400 MeV$, for the Pion, $\xi^2$ must be $2.6$, as found from Eqn.(\ref{3}). These values reproduce the experimental value of $0.43 fm^2$ for the square of the Pion charge radius.
	
	We take Case 1 ($n_r = 1$) a little further and examine other mesons as is typically done in quark models. Namely, we examine the charged Kaon by calculating the ratio of the Pion square charge radius to the Kaon square charge radius and employ the experimental value of the Kaon square charge radius of $0.31 fm^2$.  The coupling constant remains the same and so does the wavefunction. Though for the Kaon, the reduced mass $\mu = \frac{m_L m_s}{m_L + m_s}$ with $m_s$ being the strange quark mass. The result shows that there is a ratio between the light quark mass that constrains the strange quark mass to be $m_s = 1.44m_L$. This gives a strange quark mass of $576 MeV$ provided the light quark mass is $400 MeV$ -- a strange quark mass consistent with other approaches[1,2]. For the $D^{\pm}$, there is no experimental charge radius data we could find; however, there is a theoretical calculation that gives a broad range of values.  With a choice of $0.17 fm^2$ for the square of the $D$ charge radius and performing a similar ratio with the Pion as was done with the Kaon, a constraint emerges that sets the charm quark mass, $m_c = 3.88m_L$ or $m_c = 1552 MeV$ with $m_L = 400 MeV$.  Again, a charmed quark mass close to what other phenomenological models obtain. While these results are encouraging, employing Eqn.2 gives result of meson masses, that are low compare to data, or K and D as $416MeV$ and $1292MeV$ respectively. Thus, if $m_s$ and $m_c$ were adjusted away from this constrain, we can then attain the experimental values of Pion and D meson masses. Loosening this constrain would produce different charged radii; suggesting the rationale statistical approaches to meson spectroscopy. In the instance of the $B$ meson, there are certainly no experimental data nor theoretical estimates to rely on for the square of the charge radius. However, we can obtain an estimate of the bottom quark mass by employing Eqn.2 once again. The result is that  $m_b = 5539 MeV$ for a value of the bottom quark. Employing the ratio of the Pion to B meson square charge radii, yields a B square charge radius of $0.123 fm^2$. Similarly we obtain $0.107 fm^2$ for the charged T meson square charge radius.
	
	Additionally, for Case 1 ($n={(n,0)}$), we note that in the heavy quark limit where we have consistently assigned $\mu = m_L = 400 MeV$, the asymptotic value of the square charge radius is $0.107 fm^2$ or a charge radius of $0.327 fm$, which precisely with result of Pion. This result indicates scaling consistent with the QCD picture of hadrons; that is, they are not point like. It is therefore clear that within this model, the minimal charged radii of these mesons is strongly dependent on the light quark mass. We recognize that by, instead, employing the current quark value of $3.5 MeV$, as listed in the Particle Data Group Summary Table(PDT)[4], the heavy quark limit of the square charge radius is an unreasonable $1397 fm^2$. With the inclusion of confinement and$\setminus$or other more sophisticated approaches[5], this strong dependence on the light quark mass alone may be examined further.
	
	Also if the light current quark mass value of $3.5 MeV$ used from the PDT., the strange quark mass from our model would not agree with the current strange quark mass from the PDT
	
	Separately, the decay constants,
	\begin{equation}
		f_{n_r} = 2(\frac{3}{M_{n_r}})^{\frac{1}{2}}|\psi(0)|_{n_r}\footnote[4]{Reference[4]},
		\label{6}
	\end{equation}
	where $M_{n_r}$ is meson mass for all these charged pseudo scalar mesons were calculated and the results are extraordinarily larger than data; meaning that our model mesons unsurprisingly have a shorter lifetime than has been measured. Adding a confining potential such as a linear potential, that makes a large contribution to the decay rates, should fix this disparity and will also change the slopes of the potentials presented here.
	 
	The fundamental motivation for considering Case 1 is to motivate Case 2. Case 1 emerged from considering a research project created for a Master’s student (S. Lee) to determine what characteristics a simple 1/r potential could have in meson spectroscopy.  We see that while there are some simple assumptions found in Case 1, it provides useful results, namely constituent quark masses and charge radii,  that are consistent with more sophisticated models and data[5]; leading us to conclude it is not entirely unrealistic to consider only the 1/r potential. For Case 2, we ask``(1) what are the consequences or benefits of introducing numerical quantum numbers for flavor; and (2) can we reproduce the same values of charge radii for the same group of Case 1 charged pseudo scalar mesons?"
	
	So, in Case 2, we introduce a new quantum number satisfying the boundary conditions; namely, $n = n_r + n_f = (n_r, n_f) = (r, f)$ where $n_r$ is the usual radial quantum number and $n_f$ is the flavor quantum number.  As in Case 1, we assume the $\pi$, $K$, $D$, and $B$ are all in the the ground state of $n_r =1$ as before in Case 1. Since the Pion is unflavored, $(r, f) = (1,0)$; yielding identical results as in Case 1 for the Pion.  The other unequal mass states have the following assignment for $(r, f): K (1,1), D (1,2), B (1,3)$, and so forth for the $T$.  In principle, this allows for excited states of each meson (see Table \RNum{1}).  The square charge radii of Case 2 are computed from Eqn (\ref{2}) with  $<r^2>_{n} $and $n = (r, f)$.  This assignment and subsequent calculation results in $<r^2>_n = N_n a^2_0$ where, again, $a_0 = \frac{1}{\mu \xi^2}$ and $N_n$ is a number dependent on the spherical state wave function $\psi_n (r)$. The differences, between the two cases, arise in the light-heavy sector of the flavored mesons.  In Case 1, we had the flavored quark masses to consider in the usual way. In Case 2, we find that the heavy quark limit can more easily be employed even for the Kaon. The consequence of this limit in Case 2 is that the potential coupling constant significantly differs for each meson so that it still lies in the ground state of the interaction. The light quark mass remains the same for each meson in both Case 1 and 2.  Results for the phenomenological coupling constants are found in Table \RNum{3}.  All square charge radii results are identical in both Cases. From Case 2, one can determine that the mass of the first excited state for the Pion is $635MeV$; a value much lower than the $1300 MeV$ from experiment. The Pion and its excited states have the same coupling constant of $\xi^2 = 2.6$.  Also in Case 2, the excited states of the Kaon have the same coupling constant of $5.7$.  The same is true for the $D$ meson excited states; namely, they all have a coupling constant $17.2$; and so forth.
	
	In summary, results for the charge radii and light quark mass of the $\pi$, $K$, $D$, and $B$ mesons are identical in both Case 1 and Case 2 with the constant constituent light quark mass setting the scale and linking all states. 
	\begin{figure}[H]
	\centering
	\includegraphics[width=\linewidth]{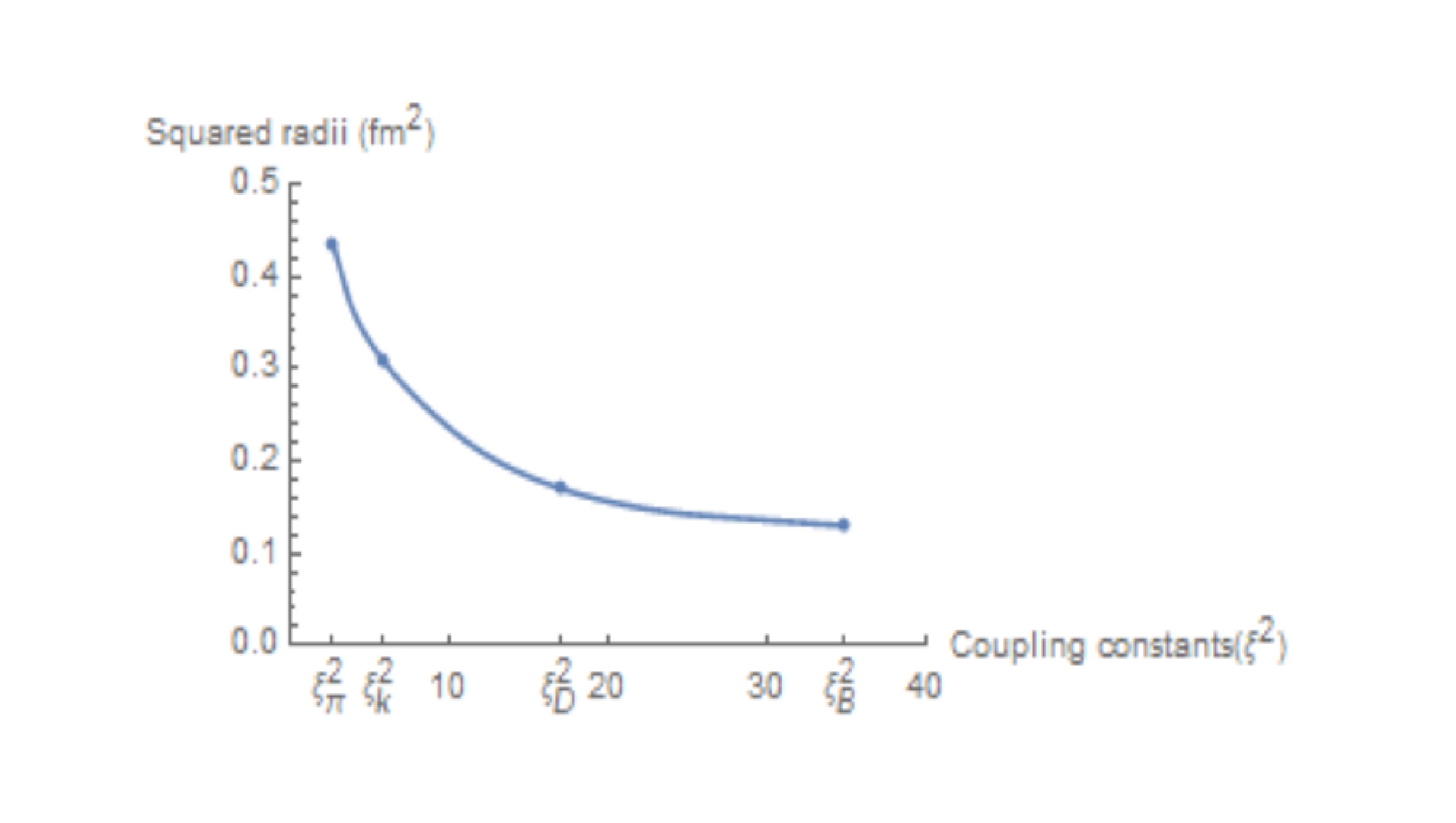}
	\caption{\small{Charged radii squared in $fm^2$ (vertical axis) vs. coupling constants (horizontal axis) in Case 2 with light quark mass of 400 MeV. Heavy quark limit is applied for Kaon and heavier mesons}}
\end{figure}
	 
	A hint of this phenomenological organization that fixes the value of the light quark can be found in earlier, covariant, elastic electromagnetic form factor calculations of the Pion and Kaon[6]; as well as the semi-leptonic decay[6] of the $K^0$.  Those calculations use a local separable contact interaction and connect to the Pion data to determine the light quark mass; which, once fixed, leads to finding the strange quark mass in the calculation of the Kaon elastic form factor; which then leads to a completely parameter free prediction of the $K^0$ semi-leptonic decay form factor, $f_+$. 
	
	Expectedly, the decay constants are very large in both Cases as there is no confining interaction to restrain decay in these models; yet, except for the Pion, there are significant differences between the two Cases. The flavor mesons decay much faster in Case 2 compared to Case 1. Also, except for the Pion, the wavefunctions in the two Cases are very different.  In Case 1, the slope for all the states is identical. In Case 2, the potential slopes are all different depending on the meson flavor state. Lastly, no heavy quark limit is taken in Case 1; while, the heavy quark limit is employed for all flavor states in Case 2.
	
	If Case 2 general features hold up in more sophisticated calculations, then it would suggest a new dynamic not before reported on.  It would be interesting to determine if one can regroup all the mesons in this way. If so, it might suggest that the gluon not only carry color as a dynamic component; but also flavor as a dynamic component.
		
	\begin{table}[H]
		\centering
		\caption{{$n_r$ (or $r$) is the radial quantum number; and $n_f$ (or $f$) is the introduced numerical flavor quantum number.
		}}
		\begin{tabular}{ c|c|c| c|c}
			\hline
			$f$ (column) & $r$(rows)& $1$ & 2& 3\\ [0.5ex]
			\hline
			0& & $\pi^+$&$\pi^{+'}$&$\pi^{+''}$\\
			\hline
			1& & $K^+$&$K^{+'}$&$K^{+''}$\\
			\hline
			2& & $D^+$&$D^{+'}$&$D^{+''}$\\
			\hline
			3& & $B^+$&$B^{+'}$&\\
			\hline
			4& & $T^+$& & \\ [0.5ex]
			\hline
		\end{tabular} 
	\end{table}
	
	\begin{table}[H]
		\centering
		\caption{}
		\begin{tabular}{ c| c|c}
			\hline
			Meson & $<r>^2 fm^2$(data)& $<r>^2 fm^2$(for n=1)\\ [0.5ex]
			\hline
			$\pi^+$ & 0.436${\cite{7}}$& 0.436\\
			
			$K^+$ & 0.31${\cite{3}}$& 0.31\\
			
			$D^+$ & 0.24\footnotemark & 0.17\\
			
			$B^+$ &$\setminus$ &0.123 \\ [0.5ex]
			\hline
		\end{tabular}		 
	\end{table}
	\footnotetext[1]{{Simple average of the results from Ref\cite{2}}}

	\begin{table}[H]
		\centering
		\caption{}
		\begin{tabular}{c|c}
			meson & $\xi^2$\\ 
			\hline
			$\pi(1,0)$& 2.6\\
			$K(1,1)$ & 5.7\\
			$D(1,2)$ & 17\\
			$B(1,3)$& 34.8\\
			\hline
		\end{tabular}
	\end{table}
	
\begin{figure}[H]
	\centering
	\includegraphics[width=\linewidth]{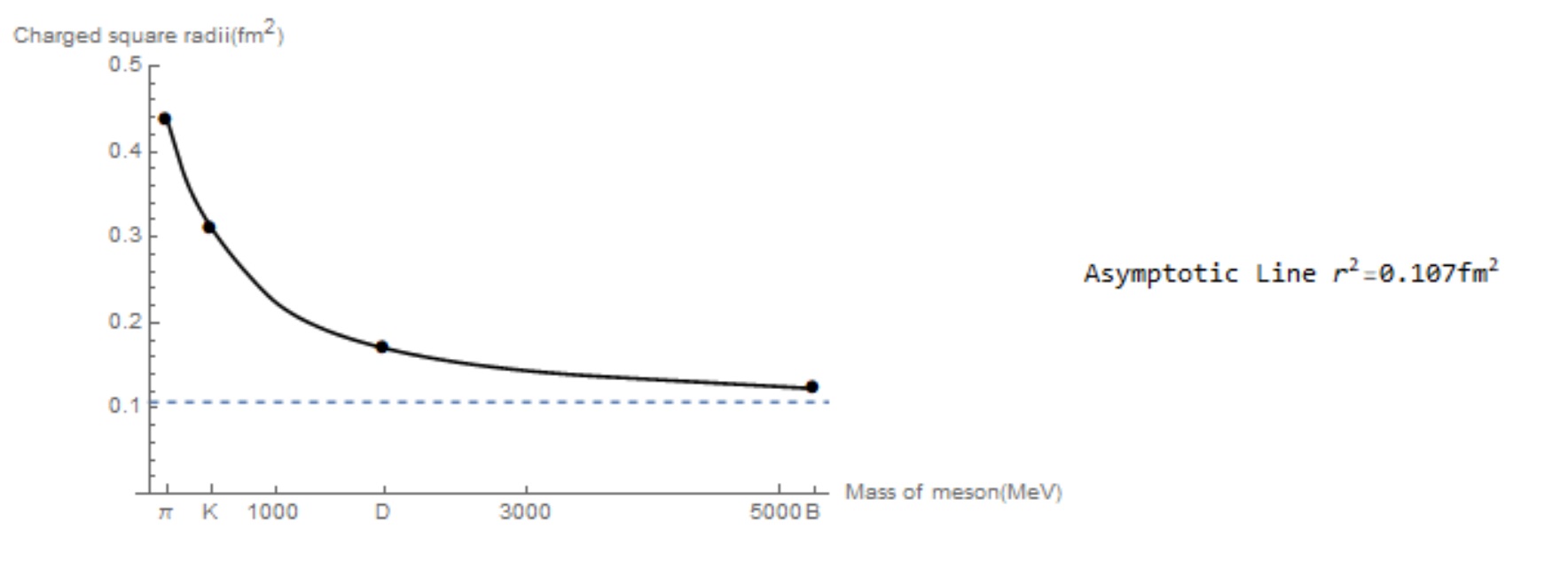}
	\caption{\small{Results of Case 1 and Case 2 are identical}}
\end{figure}

	\section*{Acknowledgements}
		We thank Ann Nelson for the review and comments of the earliest manuscript. We thank David Kaplan, Andrei Afanasev, and Franz Gross for good council and useful suggestions.  We also thank the College of William and Mary for hosting WWB during the final writing of this manuscript.

\end{document}